\documentclass[prl,twocolumn,superscriptaddress,showemail,longbibliography]{revtex4-1}
\usepackage[T1]{fontenc}
\usepackage{times}
\usepackage{units}
\usepackage{braket}
\usepackage{mathrsfs}
\usepackage{amsmath}
\usepackage{amssymb}
\usepackage{graphicx}
\usepackage{esint}
\usepackage{subfig}

\usepackage{color} 

\begin{document}


\title{{\color{black}A unified theory of spin-relaxation due to spin-orbit coupling in metals and semiconductors\\}}

\author{P\'{e}ter~Boross} \affiliation{Department of Physics, Budapest University of Technology and Economics, Budapest, Hungary} \affiliation{Department of Materials Physics, E\"{o}tv\"{o}s University, Budapest, Hungary}
\author{Bal\'{a}zs~D\'{o}ra}  \affiliation{Department of Physics, Budapest University of Technology and Economics, Budapest, Hungary} \affiliation{BME-MTA Exotic Quantum Phases Research Group, Budapest University of Technology and Economics, Budapest, Hungary}
\author{Annam\'{a}ria~Kiss} \affiliation{Department of Physics, Budapest University of Technology and Economics, Budapest, Hungary} \affiliation{Wigner Research Centre for Physics of the Hungarian Academy of Sciences, Budapest, Hungary}
\author{Ferenc~Simon} \email{simon@esr.phy.bme.hu} \affiliation{Department of Physics, Budapest University of Technology and Economics, Budapest, Hungary}


\date{\today}
\begin{abstract}
{\color{black}Spintronics is an emerging paradigm with the aim to replace conventional electronics by using electron spins as information carriers. Its utility relies on the magnitude of the spin-relaxation, which is dominated by spin-orbit coupling (SOC). Yet, SOC induced spin-relaxation in metals and semiconductors is discussed for the seemingly orthogonal cases when inversion symmetry is retained or broken by the so-called Elliott-Yafet and D'yakonov-Perel' spin-relaxation mechanisms, respectively. We unify the two theories on general grounds for a generic two-band system containing intra- and inter-band SOC. While the previously known limiting cases are recovered, we also identify parameter domains when a crossover occurs between them, i.e. when an inversion symmetry broken state evolves from a D'yakonov-Perel' to an Elliott-Yafet type of spin-relaxation and conversely for a state with inversional symmetry. This provides an ultimate link between the two mechanisms of spin-relaxation.}

\end{abstract}
\maketitle


{\color{black}A future spintronics device would perform calculations and store information using the spin-degree of
freedom of electrons with a vision to eventually replace conventional
electronics \cite{WolfSpintronics,FabianRMP,WuReview}. A spin-polarized ensemble of electrons whose spin-state
is manipulated in a transistor-like configuration and is read out
with a spin-detector (or \textit{spin-valve}) would constitute an elemental building block of a spin-transistor.} Clearly, the utility of spintronics relies on whether the
spin-polarization of the electron ensemble can be maintained sufficiently
long. The basic idea behind spintronics is that coherence of a spin-ensemble
persists longer than the coherence of electron momentum due to the
relatively weaker coupling of the spin to the environment. The coupling
is relativistic and has thus a relatively weak effect known as spin-orbit
coupling (SOC).

The time characterizing the decay of spin-polarization is the so-called
spin-relaxation time (often also referred to as spin-lattice relaxation
time), $\tau_{\text{s}}$. It can be measured either using electron
spin-resonance spectroscopy (ESR) \cite{FeherKip} or in spin-transport
experiments \cite{JohnsonSilsbeePRB1988,JedemaNat2002}. Much as the theory and experiments of spin-relaxation measurements are developed, it remains an intensively studied field for novel materials; e.g. the value of $\tau_{\text{s}}$ is the matter of intensive theoretical
studies \cite{HuertasPRB2006,FabianPRB2009a,FabianPRB2009b,CastroNetoGrapheneSO,DoraEPL2010,WuNJP2012,CastroNetoGuineaPRL2012}
and spin-transport experiments \cite{TombrosNat2007,KawakamiPRL2010,KawakamiBilayer,GuntherodtBilayer}
in graphene at present.

\begin{figure}
\begin{centering}
\includegraphics[width=0.9\columnwidth]{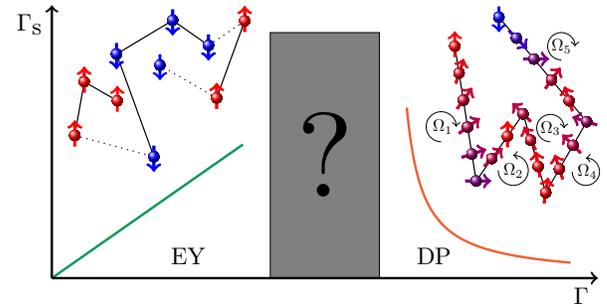}
\par\end{centering}
\caption{Schematics of the Elliott-Yafet and the D'yakonov-Perel' mechanisms: $\Gamma_{\text{s}}\propto \Gamma$ in the EY scenario and spin-scattering occurs rarely (typically for every $10^{4}..10^{6}$th momentum scattering
in alkali metals), whereas the spin direction continuously precesses
around the internal magnetic field due to SOC in the DP scenario, resulting in $\Gamma_{\text{s}}\propto 1/\Gamma$. It is the topic of the present paper, how these two distinct regimes are related to each other.}
\label{SpinRelaxSchematics}
\end{figure}

The two most important spin-relaxation mechanisms in metals and semiconductors are
the so-called Elliott-Yafet (EY) and the D'yakonov-Perel' (DP) mechanisms. These are conventionally
discussed along disjoint avenues, due to reasons described below. Although the interplay between these mechanisms has been studied in semiconductors \cite{WuReview,PikusTitkovBookChapter,AverkievGolubWillander,GlazovShermanDugaev}, no attempts have been made to unify their descriptions. We note that a number of other spin-relaxation mechanisms, e.g. that involving nuclear-hyperfine interaction, are known \cite{FabianRMP,WuReview}.

The EY theory \cite{Elliott,YafetPL1983} describes spin-relaxation
in metals and semiconductors with inversion symmetry. Therein, the
SOC does not split the spin-up/down states ($\Ket{\uparrow}$,
$\Ket{\downarrow} $) in the conduction band \cite{Endnote},
however the presence of a near lying band weakly mixes these states
while maintaining the energy degeneracy. The nominally up state
reads: $\Ket{\tilde{\uparrow}} =a_{k}\Ket{\uparrow} +b_{k}\Ket{\downarrow} $
(here $a_{k},\, b_{k}$ are band structure dependent) and $b_{k}/a_{k}=L/\Delta$,
where $L$ is the SOC matrix element between the adjacent bands and
$\Delta$ is their separation. E.g.~in alkali metals $L/\Delta\approx10^{-2}..10^{-3}$
[Ref.~\cite{YafetPL1983}]. Elliott showed using first order time-dependent
perturbation theory that an electron can flip its spin with probability
$\left(L/\Delta\right)^{2}$ at a momentum scattering event.
As a result, the spin scattering rate ($\Gamma_{\textrm{s}}=\hbar/2\tau_{\textrm{s}}$)
reads:

\begin{equation}
\Gamma_{\text{s,EY}}\approx\left(\frac{L}{\Delta}\right)^{2}\Gamma,
\end{equation}

\noindent where $\Gamma=\hbar/2\tau$ is the quasi-particle scattering
rate with $\tau$ being the corresponding momentum scattering (or
relaxation) time. This mechanism is schematically depicted in Fig.~\ref{SpinRelaxSchematics}a.

For semiconductors with zinc-blende crystal structure, such as e.g. GaAs, the lack of inversion symmetry results
in an efficient relaxation mechanism, the D'yakonov-Perel' spin-relaxation
\cite{DyakonovPerelSPSS1972}. Therein, the spin-up/down energy levels
in the conduction bands are split. The splitting acts on the electrons
as if an internal, $k$-dependent magnetic field would be present,
around which the electron spins precess with a Larmor frequency of
$\Omega(k)=\mathscr{L}(k)/\hbar$. Here $\mathscr{L}(k)$ is the energy
scale for the inversion symmetry breaking induced SOC. Were no momentum
scattering present, the electron energies would acquire a distribution
according to $\hbar\Omega(k)$. In the presence of momentum scattering
which satisfies $\Omega(k)\cdot \tau\ll1$, the distribution is \textquotedbl{}motionally-narrowed\textquotedbl{}
and the resulting spin-relaxation rate reads:
\begin{equation}
\Gamma_{\text{s,DP}}\approx\frac{\mathscr{L}^{2}}{\Gamma}.
\end{equation}
\noindent This situation is depicted in Fig.~\ref{SpinRelaxSchematics}b.
Clearly, the EY and DP mechanisms result in different dependence
on $\Gamma$ which is often used for the empirical assignment
of the relaxation mechanism \cite{TombrosPRL2008}. {\color{black}}

The observation of an anomalous temperature dependence of the spin-relaxation
time in MgB$_{2}$ \cite{SimonPRL2008} and the alkali fullerides
\cite{DoraPRL2009} and the development of a generalization of the
EY theory highlighted that the spin-relaxation theory is not yet complete.
In particular, the first order perturbation theory of Elliott breaks
down when the quasi-particle scattering rate is not negligible compared
to the other energy scales. One expects similar surprises for the
DP theory when the magnitude of e.g.~the Zeeman energy is considered
in comparison to the other relevant energy scales.

Herein, we develop a general and robust theory of spin-relaxation
in metals and semiconductors including SOC between different bands
and the same bands, provided the crystal symmetry allows for the latter.
We employ the Mori-Kawasaki theory which considers the kinetic motion
of the electrons under the perturbation of the SOC. We obtain a general
result which contains both the EY and the DP mechanisms as limits
when the quasi-particle scattering and the magnetic field are small.
Interesting links are recognized between the two mechanisms when these
conditions are violated: the EY mechanism appears to the DP-like when
$\Gamma$ is large compared to $\Delta$ and the DP mechanism appears
to be EY-like when the Zeeman energy is larger than $\Gamma$. Qualitative
explanations are provided for these analytically observed behaviors.


\section*{Results} 

The minimal model of spin-relaxation is a
four-state (two bands with spin) model Hamiltonian for a two-dimensional electron gas (2DEG) in a  magnetic field, which reads:

\begin{subequations}
\begin{align}
\mathcal{H} & =\mathcal{H}_{\textrm{0}}+\mathcal{H}_{\textrm{Z}}+\mathcal{H}_{\textrm{scatt}}+\mathcal{H}_{\textrm{SO}}\\
\mathcal{H}_{\textrm{0}} & =\sum_{k,\alpha,s}\epsilon_{k,\alpha}\, c_{k,\alpha,s}^{\dagger}c_{k,\alpha,s}^{}\\
\mathcal{H}_{\textrm{Z}} & =\Delta_{\textrm{Z}}\sum_{k,\alpha,s}s\, c_{k,\alpha,s}^{\dagger}c_{k,\alpha,s}^{}\\
\mathcal{H}_{\textrm{SO}} & =\sum_{k,\alpha,\alpha',s,s'}L_{\alpha,\alpha',s,s'}\left(k\right)\, c_{k,\alpha,s}^{\dagger}c_{k,\alpha',s'}^{},
\end{align}
\end{subequations}
where $\alpha=1$ (nearby), $2$ (conduction) is the
band index with $s=\left(\uparrow\right),\left(\downarrow\right)$
spin, $\epsilon_{k,\alpha}=\hbar^{2}k^{2}/2m_{\alpha}^{*}-\delta_{\alpha,1}\Delta$ is the single-particle dispersion
with $m_{\alpha}^{*}=(-1)^{\alpha}m^{*}$ effective mass and $\Delta$ band gap,
$\Delta_{Z}=g\mu_{\textrm{B}}B_{z}$ is the Zeeman energy. $\mathcal{H}_{\textrm{scatt}}$
is responsible for the finite quasi-particle lifetime due to impurity and electron-phonon scattering and $L_{\alpha,\alpha',s,s'}\left(k\right)$
is the SOC.

\begin{figure}
\begin{centering}
\includegraphics[width=0.85\columnwidth]{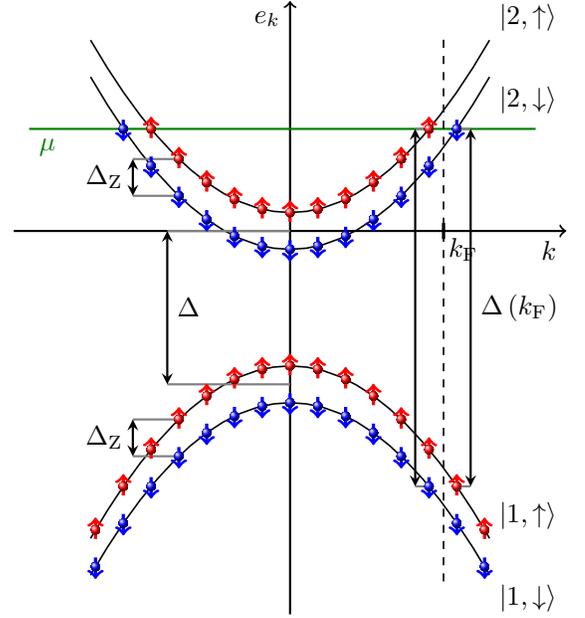}
\par\end{centering}
\caption{\textit{Color online.} The band structure of a 2DEG in a magnetic
field. The effects of the weak SOC are not shown. Vertical arrows show the energy separations between the relevant bands.}
\label{BandStructure}
\end{figure}

The corresponding band structure is depicted in Fig.~\ref{BandStructure}. The eigenenergies and eigenstates without SOC are

\begin{subequations}
\begin{align}
e_{k,\alpha,s} & =\epsilon_{k,\alpha}+s\Delta_{\textrm{Z}}\\
\left|1,\downarrow\right\rangle  & =\left[1,0,0,0\right]^{\intercal}\quad\left|1,\uparrow\right\rangle =\left[0,1,0,0\right]^{\intercal}\\
\left|2,\downarrow\right\rangle  & =\left[0,0,1,0\right]^{\intercal}\quad\left|2,\uparrow\right\rangle =\left[0,0,0,1\right]^{\intercal}.
\end{align}
\end{subequations}

The most general expression of the SOC for the above levels reads:

\begin{align}
L_{\alpha,\alpha',s,s'}\left(k\right) & =\left(\begin{array}{cccc}
\mathscr{L}_{\uparrow\uparrow} & \mathscr{L}_{\downarrow\uparrow} & L_{\uparrow\uparrow} & L_{\downarrow\uparrow}\\
\mathscr{L}_{\uparrow\downarrow} & \mathscr{L}_{\downarrow\downarrow} & L_{\uparrow\downarrow} & L_{\downarrow\downarrow}\\
L_{\uparrow\uparrow} & L_{\downarrow\uparrow} & \mathscr{L}_{\uparrow\uparrow} & \mathscr{L}_{\downarrow\uparrow}\\
L_{\uparrow\downarrow} & L_{\downarrow\downarrow} & \mathscr{L}_{\uparrow\downarrow} & \mathscr{L}_{\downarrow\downarrow}
\end{array}\right),
\end{align}
where $\mathscr{L}_{ss'}\left(k\right)$, $L_{ss'}\left(k\right)$ are
the wavevector dependent intra- and inter-band terms, respectively, which are phenomenological, i.e. not related to a microscopic model. The terms
mixing the same spin direction can be ignored as they commute with the $S_{z}$ operator and do not cause spin-relaxation. The SOC terms contributing to spin-relaxation are
\begin{align}
L_{\alpha,\alpha',s,s'}\left(k\right) & =\left(\begin{array}{cccc}
0 & \mathscr{L} & 0 & L\\
\mathscr{L}^{\dagger} & 0 & L^{\dagger} & 0\\
0 & L & 0 & \mathscr{L}\\
L^{\dagger} & 0 & \mathscr{L}^{\dagger} & 0
\end{array}\right).
\end{align}

\begin{table}
\begin{centering}
\begin{tabular*}{1\columnwidth}{@{\extracolsep{\fill}}ccc}
\hline \hline
 & inversion symmetry & broken inv. symm.\tabularnewline
\hline
$\mathscr{L}$ & $0$ & $ $finite\tabularnewline
$L$ & finite & finite\tabularnewline
$\varepsilon_{k,\uparrow}-\varepsilon_{k,\downarrow}$ & 0 & finite\tabularnewline
\hline \hline
\end{tabular*}
\par\end{centering}
\begin{footnotesize}
\begin{raggedright}
\begin{tabbing}
\end{tabbing}
\par\end{raggedright}
\end{footnotesize}
\caption{Effect of the presence or absence of the inversion symmetry on the intra- ($\mathscr{L}$) and inter-band ($L$) SOC and on the energy splitting of spin-states in the same band, $\varepsilon_{k,\uparrow}-\varepsilon_{k,\downarrow}$.}
\label{SOCsymmtable}
\end{table}

Table \ref{SOCsymmtable}.~summarizes the role of the inversion symmetry on the SOC parameters. For a material with inversion symmetry, the Kramers theorem dictates (without magnetic field) that $\epsilon_{\uparrow}\left(k\right)=\epsilon_{\downarrow}\left(k\right)$ and thus $\mathscr{L}=0$, which term would otherwise split the spin degeneracy in the same band. When the inversion symmetry is broken, $\mathscr{L}$ is finite and the previous degeneracy is reduced to a weaker condition: $\epsilon_{\uparrow}\left(k\right)=\epsilon_{\downarrow}\left(-k\right)$ as the time-reversal symmetry is retained.

We consider the SOC as the smallest energy scale in our model ($\mathscr{L}\left(k_{\textrm{F}}\right),L\left(k_{\textrm{F}}\right))$,
while we allow for a competition of the other energy scales, namely $\Delta_{\textrm{Z}}$, $\Gamma$ and $\Delta$, which can be of the same order of magnitude, as
opposed to the conventional EY or DP case.
We are mainly interested in the regime of a weak SOC, moderate magnetic
fields, high occupation, and a large band gap. We treat the quasi-particle scattering rate to infinite order thus large values of $\Gamma$ are possible.

The energy spectrum of the spins (or the ESR line-width) can be calculated from the Mori-Kawasaki formula
\cite{PTP.28.971,PhysRevB.65.134410}, which relies on the assumption
that the line-shape is Lorentzian.
This was originally proposed for localized spins (e.g.~Heisenberg-type models) but it can be extended to itinerant electrons.
The standard (Faraday) ESR configuration measures the absorption
of the electromagnetic wave polarized perpendicular to the static
magnetic field. The ESR signal intensity is
\begin{align}
I\left(\omega\right) & =\frac{B_{\perp}^{2}\omega}{2\mu_{0}}\chi_{\perp}''\left(q=0,\omega\right)V,
\end{align}
where $B_{\perp}$ is the magnetic induction of the electromagnetic
radiation, $\chi_{\perp}''$ is the imaginary part of the spin-susceptibility,
$\mu_{0}$ is the permeability of vacuum, and $V$ is the sample volume.
The spin-susceptibility is related to the retarded Green's function as
\begin{align}
\chi_{\perp}''\left(\omega\right) & =-\textrm{Im}G_{S^+S^-}^{\textrm{R}}\left(\omega\right),
\end{align}
with $S^{\pm}=S_{x}\pm iS_{y}$, from which the ESR spectrum
can be obtained.

The equation of motion of the $S^{+}$ operator
reads as
\begin{align}
\frac{dS^{+}}{dt} & =\frac{i}{\hbar}\left[\mathcal{H},S^{+}\right]=\underset{-i\Delta_{Z}\frac{S^{+}}{\hbar}}{\underbrace{\frac{i}{\hbar}\left[\mathcal{H}_{\textrm{Z}},S^{+}\right]}}+\underset{i\mathcal{A}}{\underbrace{\frac{i}{\hbar}\left[\mathcal{H}_{\textrm{SO}},S^{+}\right]}},
\end{align}
where $\mathcal{A}=\frac{1}{\hbar}\left[\mathcal{H}_{\textrm{SO}},S^{+}\right]$
is the consequence of the SOC. The Green's function of $S^{+}S^{-}$
is obtained from the Green's function of $\mathcal{A}^{\dagger}\mathcal{A}$
as
\begin{align}
G_{S^{+}S^{-}}^{\textrm{R}}\left(\omega\right) & =\frac{2\left\langle S_{z}\right\rangle }{\omega-\frac{\Delta_{Z}}{\hbar}}+\frac{-\left\langle \left[\mathcal{A}\left(0\right),S^{-}\left(0\right)\right]\right\rangle +G_{\mathcal{A}^{\dagger}\mathcal{A}}^{\textrm{R}}\left(\omega\right)}{\left(\omega-\frac{\Delta_{Z}}{\hbar}\right)^{2}}.
\end{align}
The second term is zero without SOC thus a completely sharp resonance occurs at the Zeeman energy. The line-shape is Lorentzian for a weak SOC:
\begin{align}
G_{S^{+}S^{-}}^{\textrm{R}}\left(\omega\right) & = \frac{2\hbar\langle S_z\rangle}{\hbar\omega-\Delta_Z-\Sigma(\omega)},
\end{align}
where the self-energy is
\begin{align}
\Sigma\left(\omega\right) & =\frac{-\left\langle \left[\mathcal{A}\left(0\right),S^{-}\left(0\right)\right]\right\rangle +G_{\mathcal{A}^{\dagger}\mathcal{A}}^{\textrm{R}}\left(\omega\right)}{2\left\langle S_{z}\right\rangle },
\end{align}
which is assumed to be a smooth function of $\omega$ near $\Delta_{Z}/\hbar$.

The spin-relaxation rate is equal to the imaginary part of $\Sigma\left(\omega\right)$
as
\begin{align}
\Gamma_{\textrm{s}} & =\frac{\textrm{Im}G_{\mathcal{A}^{\dagger}\mathcal{A}}^{\textrm{R}}\left(\frac{\Delta_{Z}}{\hbar}\right)}{2\left\langle S_{z}\right\rangle }.
\label{gammas}
\end{align}

The $G_{\mathcal{A}^{\dagger}\mathcal{A}}^{\textrm{R}}\left(\omega\right)$ correlator
is obtained from the Matsubara Green's
function of $\mathcal{A}^{\dagger}\mathcal{A}$, given by
\begin{align}
\mathcal{G}_{\mathcal{A}^{\dagger}\mathcal{A}}\left(i\nu_{m}\right)=\int_{0}^{\beta\hbar}d\tau e^{i\nu_{m}\tau}\left\langle \mathcal{T}_{\tau}\mathcal{A}^{\dagger}\left(\tau\right)\mathcal{A}\left(0\right)\right\rangle.
\label{a+a}
\end{align}
The effect of $\mathcal{H}_{\textrm{scatt}}$
is taken into account in the Green's function by a finite, constant momentum-scattering rate.

The most compact form of the spin-relaxation is obtained when the Fermi energy is not close to the bottom of the conduction band ($\mu\gtrsim\triangle$) and a calculation (detailed in the {\color{black}Methods section}) using Eq.~\eqref{gammas} leads to our main result:

\begin{align}
\Gamma_{\textrm{s}} & =\frac{4\Gamma\left|\mathscr{L}\left(k_{\textrm{F}}\right)\right|^{2}}{4\Gamma^{2}
+\Delta_{\textrm{Z}}^{2}}+\frac{4\Gamma\left|L\left(k_{\textrm{F}}\right)\right|^{2}}{4\Gamma^{2}+\Delta^{2}\left(k_{\textrm{F}}\right)},
\label{vegsogammas}
\end{align}

\noindent Results in more general cases are discussed in the Supplementary Material.

\section*{Discussion}

\begin{figure}
\begin{centering}
\includegraphics[width=0.85\columnwidth]{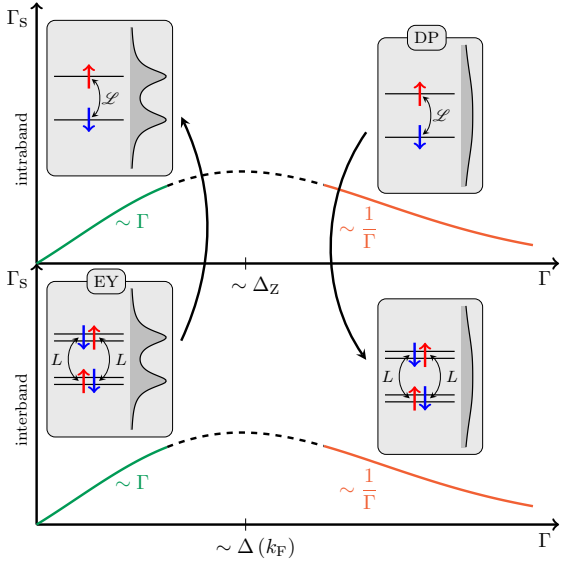}
\par\end{centering}
\caption{$\Gamma_{\textrm{s}}$ as a function of $\Gamma$ according to Eq. \eqref{vegsogammas} concerning separately the contributions due to intra- (upper) and inter-band (lower) SOC's. The insets are schematics of the band-structure including the broadening due to $\Gamma$ (double arrows indicate the matrix elements). The arrows indicate the equivalence of the different spin-relaxation regimes. The conventional DP and EY scenarios are those in the top-right and bottom-left corners, respectively.}
\label{EYDPSymm}
\end{figure}

 {\color{black}According to Eq. \eqref{vegsogammas}, the} contributions from intra- ($\mathscr{L}\left(k_{\textrm{F}}\right)$) and inter-band ($L\left(k_{\textrm{F}}\right)$) processes are additive to lowest order in the SOC and have a surprisingly similar form. A competition is observed between lifetime induced broadening (due to $\Gamma$) and the energy separation between states ($\Delta\left(k_{\textrm{F}}\right)$ or $\Delta_{\textrm{Z}}$). The situation, together with schematics of the corresponding
band-structures, is shown in Fig.~\ref{EYDPSymm}. When the broadening is much smaller than the energy separation, the relaxation is EY-like, $\Gamma_{\textrm{s}}\propto \Gamma$, even when the intra-band SOC dominates, i.e.~for a material with inversion symmetry breaking. This situation was also studied in Ref.~\cite{Ivchenko,BalentsBurkovPRB2004} and it may be realized in III-V semiconductors in high magnetic fields. For metals with inversion symmetry, this is the canonical EY regime.

When the states are broadened beyond distinguishability (i.e.~$\Gamma\gg\Delta\left(k_{\textrm{F}}\right)$ or $\Delta_{\textrm{Z}}$), spin-relaxation is caused by two quasi-degenerate states and the relaxation is of DP-type, $\Gamma_{\textrm{s}}\propto 1/\Gamma$, even for a metal with inversion symmetry, $\mathscr{L}=0$. For usual metals, the $\Gamma\gg\Delta\left(k_{\textrm{F}}\right)$, criterion implies a breakdown of the quasi-particle picture as therein $\Delta\left(k_{\textrm{F}}\right)$ is comparable to the bandwidth, thus this criterion means strong-localization. In contrast, metals with nearly degenerate bands remain metallic as e.g. MgB$_2$ (Ref.~\cite{SimonPRL2008}) and the alkali fullerides (K$_3$C$_{60}$ and Rb$_3$C$_{60}$) (Ref.~\cite{DoraPRL2009}), which are strongly correlated metals with large $\Gamma$. When the intra-band SOC dominates, i.e.~for a strong inversion symmetry breaking, this is the canonical DP regime. {\color{black} These observations provide the ultimate link between these two spin-relaxation mechanisms, which are conventionally thought as being mutually exclusive.}

Similar behavior can be observed in other models (details are given in the Supplementary material), $\Gamma_{\textrm{s}}\propto \Gamma$ and $\Gamma_{\textrm{s}}\propto 1/\Gamma$ remain valid in the two different limits but the intermediate behavior is not universal. A particularly compelling situation is the case of graphene where a four-fold degeneracy is present at the Dirac-point and both inter- and intra-band SOC are present thus changing the chemical potential would allow to map the crossovers predicted herein.

\section*{Methods}
{\color{black}

We consider Eq.~(14) as a starting point. The Matsubara Green's function of $\mathcal{A}^{\dagger}\mathcal{A}$ can be written as
\begin{widetext}
\begin{align}
\mathcal{G}_{\mathcal{A}^{\dagger}\mathcal{A}}\left(i\nu_{m}\right)
=\frac{1}{\beta\hbar}\underset{i\omega_{n},k,\alpha,\alpha',s,s'}{\sum}\bigl|\bigl\langle k,\alpha,s\bigl|\mathcal{A}^{\dagger}\bigr|k,\alpha',s'\bigr\rangle\bigr|^{2}
\mathcal{G}_{\alpha,s}\left(i\omega_{n},k\right)\mathcal{G}_{\alpha',s'}\left(i\omega_{n}+i\nu_{m},k\right),
\label{a+a}
\end{align}
\end{widetext}
where
\begin{align}
\mathcal{G}_{\alpha,s}\left(i\omega_{n},k\right) & =\frac{1}{i\omega_{n}-\frac{1}{\hbar}\left(e_{k,\alpha,s}-\mu\right)+i\frac{\Gamma}{\hbar}\textrm{sgn}\left(\omega_{n}\right)}
\end{align}
is the Matsubara Green's function of fermionic field operators in band ($\alpha$) and spin ($s$).
The effect of $\mathcal{H}_{\textrm{scatt}}$
is taken into account by the finite momentum-scattering rate, $\Gamma$.

Using
the relationship between the Green's function and spectral density, the Matsubara summation in Eq.~\eqref{a+a} yields
\begin{widetext}
\begin{align}
\mathcal{G}_{\mathcal{A}^{\dagger}\mathcal{A}}\left(i\nu_{m}\right)=\frac{1}{4\pi^{2}}\underset{k,\alpha,\alpha',s,s'}{\sum}\bigl|\bigl\langle k,\alpha,s\bigl|
\mathcal{A}^{\dagger}\bigr|k,\alpha',s'\bigr\rangle\bigr|^{2}\times\nonumber\\
\times\int_{-\infty}^{\infty}\int_{-\infty}^{\infty}d\omega'd\omega''\,\frac{n_{\textrm{F}}\left(\omega''\right)-n_{\textrm{F}}\left(\omega'\right)}{i\nu_{m}-\omega''+\omega'}
\rho_{\alpha,s}\left(\omega',k\right)\rho_{\alpha',s'}\left(\omega'',k\right),
\end{align}
\end{widetext}
where
\begin{align}
\rho_{\alpha,s}\left(\omega,k\right) & =\frac{-2\frac{\Gamma}{\hbar}}{\left[\omega-\frac{1}{\hbar}\left(e_{k,\alpha,s}-\mu\right)\right]^{2}+\left(\frac{\Gamma}{\hbar}\right)^{2}}
\end{align}
is the spectral density. By taking the imaginary part after analytical continuation, the energy integrals can be calculated at zero temperature.
Then, by replacing momentum summation with integration, we obtain
\begin{align}
\textrm{Im}G_{\mathcal{A}^{\dagger}\mathcal{A}}^{\textrm{R}}\left(\omega\right)=\frac{A}{8\pi^{2}}\int_{0}^{\infty}dk\,k
\underset{\alpha,\alpha',s,s'}{\sum}\bigl|\bigl\langle k,\alpha,s\bigl|\mathcal{A}^{\dagger}\bigr|k,\alpha',s'\bigr\rangle\bigr|^{2}\xi_{\alpha,s,\alpha',s'}
\left(k,\omega\right),
\end{align}
where
\begin{multline}
\xi_{\alpha,s,\alpha',s'}\left(k,\omega\right)=\frac{4\hbar\Gamma}{4\Gamma^{2}+\left(\tilde{e}_{k,\alpha,s}-\tilde{e}_{k,\alpha',s'}-\hbar\omega\right)^{2}}\times\\
\times\left[\arctan\left(\frac{\tilde{e}_{k,\alpha,s}}{\Gamma}\right)-\arctan\left(\frac{\tilde{e}_{k,\alpha',s'}}{\Gamma}\right)-\arctan\left(\frac{\tilde{e}_{k,\alpha,s}-\hbar\omega}{\Gamma}\right)+\arctan\left(\frac{\tilde{e}_{k,\alpha',s'}+\hbar\omega}{\Gamma}\right)\right]+\\
+\frac{4\hbar\Gamma^{2}}{\left(\tilde{e}_{k,\alpha,s}-\tilde{e}_{k,\alpha',s'}-\hbar\omega\right)\left[4\Gamma^{2}+\left(\tilde{e}_{k,\alpha,s}-\tilde{e}_{k,\alpha',s'}-\hbar\omega\right)^{2}\right]}\times\nonumber\\
\times
\ln\frac{\left[\tilde{e}_{k,\alpha,s}^{2}+\Gamma^{2}\right]\left[\tilde{e}_{k,\alpha',s'}^{2}+\Gamma^{2}\right]}{\left[\left(\tilde{e}_{k,\alpha,s}-\hbar\omega\right)^{2}+\Gamma^{2}\right]\left[\left(\tilde{e}_{k,\alpha',s'}+\hbar\omega\right)^{2}+\Gamma^{2}\right]}
\end{multline}
and $\tilde{e}_{k,\alpha,s}=e_{k,\alpha,s}-\mu$,  $A$
is the area of the 2DEG. The matrix elements of the $\mathcal{A}^{\dagger}$ operator
are
\begin{align}
\bigl\langle k,\alpha,s\bigl|\mathcal{A}^{\dagger}\bigr|k,\alpha',s'\bigr\rangle=\left(\begin{array}{cccc}
-\mathscr{L} & 0 & -L & 0\\
0 & \mathscr{L} & 0 & L\\
-L & 0 & -\mathscr{L} & 0\\
0 & L & 0 & \mathscr{L}
\end{array}\right).
\end{align}

We determine the expectation value of the $z$-component of electron spin following similar steps as
\begin{widetext}
\begin{multline}
\left\langle S_{z}\right\rangle =\underset{i\omega_{n},k,\alpha,s}{\sum}\bigl\langle k,\alpha,s\bigl|S_{z}
\bigr|k,\alpha,s\bigr\rangle\mathcal{G}_{\alpha,s}\left(i\omega_{n},k\right)=\\
=\underset{k,\alpha,s}{\sum}\bigl\langle k,\alpha,s\bigl|S_{z}\bigr|k,\alpha,s\bigr\rangle\int_{-\infty}^{\infty}
\frac{d\omega'}{2\pi}\,n_{\textrm{F}}\left(\omega'\right)\rho_{\alpha,s}\left(\omega',k\right)
=\underset{k,\alpha,s}{\sum}\bigl\langle k,\alpha,s\bigl|S_{z}\bigr|k,\alpha,s\bigr\rangle\zeta_{\alpha,s}\left(k\right),
\end{multline}
\end{widetext}

where
\begin{widetext}
\begin{align}
\zeta_{\alpha,s}\left(k\right) & =\frac{1}{2}-\frac{1}{\pi}\arctan\left(\frac{\tilde{e}_{k,\alpha,s}}{\Gamma}\right).
\end{align}
\end{widetext}

The matrix elements of the $S_{z}$ operator are
\begin{align}
\bigl\langle k,\alpha,s\bigl|S_{z}\bigr|k,\alpha,s\bigr\rangle & =\frac{\hbar}{2}\left(\begin{array}{cccc}
-1 & 0 & 0 & 0\\
0 & 1 & 0 & 0\\
0 & 0 & -1 & 0\\
0 & 0 & 0 & 1
\end{array}\right).
\end{align}
The spin-relaxation rate can be obtained as
\begin{align}
\Gamma_{\textrm{s}} & =\frac{\textrm{Im}G_{\mathcal{A}^{\dagger}\mathcal{A}}^{\textrm{R}}\left(\frac{\Delta_{Z}}{\hbar}\right)}{2\left\langle S_{z}\right\rangle } = \Gamma_{\textrm{s}}^{\textrm{intra}}+\Gamma_{\textrm{s}}^{\textrm{inter}}.
\end{align}
We note this is the sum of intra- and inter-band terms which are described separately.
}

\section*{Acknowledgements}

We thank A. P\'{a}lyi for enlightening discussions. Work supported by the ERC Grant Nr. ERC-259374-Sylo, the Hungarian Scientific
Research Funds Nos.  K72613, K73361,
K101244, PD100373, the New Sz\'{e}chenyi Plan Nr. T\'{A}MOP-4.2.2.B-10/1.2010-0009, and by the Marie Curie
Grants PIRG-GA-2010-276834. BD acknowledges the Bolyai Program of the Hungarian Academy of Sciences.

{\color{black}
\section*{Author contributions}
PB carried out all calculations under the guidance of BD. AK contributed to the discussion and FS initiated the development of the unified theory.
All authors contributed to the writing of the manuscript
}

\begin{widetext}

\section*{SUPPLEMENTARY INFORMATION}

This {\color{black}Supplementary Material} is organized as follows: we discuss the generalization of the spin-relaxation for two kinds of band dispersions (quadratic and linear or linearized) when the restrictions concerning the relative magnitude of the parameters ($\Delta, \mu, m^{*}$) are lifted. We arrive at the overall conclusion that while the quantitative details of the $\Gamma$ dependent spin-relaxation ($\Gamma_\textrm{s}$) are modified and no closed form of the result can be provided in the most general case, the overall trends, which characterize the EY and DP behaviors and in particular the crossover between the two, remain valid. Wherever possible, we provide closed form results though. {\color{black}Finally,} we discuss the spin-relaxation for a model where only Rashba-type spin-relaxation is present.

\section{Spin-relaxation for different model dispersions}

\subsection{Quadratic dispersion model}

\begin{figure}
\begin{centering}
\includegraphics[width=0.45\columnwidth]{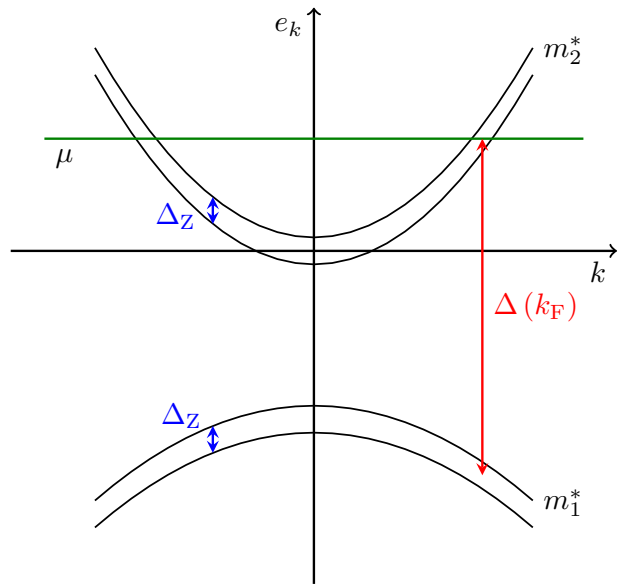}
\par\end{centering}
\caption{Band structure of quadratic dispersion model. Vertical arrows show the energy separations between the relevant bands.}
\label{QuadDisp}
\end{figure}

First, we discuss a quadratic model with the $\epsilon_{k,\alpha}=\hbar^{2}k^{2}/2m_{\alpha}^{*}-\delta_{\alpha,1}\Delta$ single-particle dispersion. In the conduction band, the quasi-particles are electron-type (i.e.~$ m_{2}^{*}>0 $) however the quasi-particles of a nearby band are hole-type (i.e.~$ m_{2}^{*} <0 $). The band structure is depicted in Fig.~\ref{QuadDisp}. This model describes well two bands of the spectrum of semiconductors, however {\color{black}in a realistic case (e.g.~for Si and GaAS)} there are {\color{black}more} nearby bands characterized by different band gaps and effective masses.

\subsubsection{{\color{black}The} intra-band term}
An important and general limit of the model is when the Zeemann energy is much smaller than the band gap (i.e.~$\Delta_{\textrm{Z}} \ll \Delta$), and {\color{black} both the} spin-up and spin-down states are occupied in the conduction band (i.e.~$\Delta_{\textrm{Z}} \ll \mu$). In this limit, the intra-band term can be expressed as
\begin{align}
\Gamma_{\textrm{s}}^{\textrm{intra}}=\frac{4\left|\mathscr{L}\left(k_{\textrm{F}}\right)\right|^{2}\Gamma}{4\Gamma^{2}+\Delta_{\textrm{Z}}^{2}}.
\end{align}
This term comes from processes {\color{black} within} the conduction band and the nearby band does not give a contribution to the intra-band term.

\subsubsection{{\color{black}The} inter-band term in the general case}
In the $\Delta_{\textrm{Z}} \ll \mu ,\Delta$ limit, {\color{black}the inter-band term can be determined but it takes a more complicated form}. When the broadening is much smaller than the energy separation at the Fermi wavenumber (i.e.~$\Gamma \ll    \Delta\left(k_{\textrm{F}}\right)=\Delta+\left(1-m_{2}^{*}/m_{1}^{*}\right)\mu$) the inter-band spin-relaxation has the form of
\begin{align}
\Gamma_{\textrm{s}}^{\textrm{inter}}=\frac{4\left|L\left(k_{\textrm{F}}\right)\right|^{2}}{\Delta^{2}\left(k_{\textrm{F}}\right)}\Gamma,
\end{align}
which is directly proportional to the momentum-scattering rate.

When the broadening is much larger than the energy separation (i.e.~$\Gamma \gg \Delta\left(k_{\textrm{F}}\right)$) the spin-relaxation reads
\begin{align}
\Gamma_{\textrm{s}}^{\textrm{inter}}=\frac{-4m_{1}^{*}m_{2}^{*}\left|L\left(k_{\textrm{F}}\right)\right|^{2}}{\left(m_{2}^{*}-m_{1}^{*}\right)^{2}}\frac{1}{\Gamma},
\end{align}
which is inversely proportional to the momentum-scattering rate.


\subsubsection{{\color{black}The} inter-band term in the case of $m_{1}^{*}=-m_{2}^{*}$}
If the effective masses in the two bands {\color{black} have different signs but the same magnitude}, the spin-relaxation rate is obtained as
\begin{align}
\Gamma_{\textrm{s}}^{\textrm{inter}}=\frac{4\left|L\left(k_{\textrm{F}}\right)\right|^{2}\Gamma}{4\Gamma^{2}+\Delta^{2}\left(k_{\textrm{F}}\right)}\left[1+\frac{\Gamma\ln\frac{\Gamma^{2}+\mu^{2}}{\Gamma^{2}+\left(\Delta+\mu\right)^{2}}}{\left(\Delta+2\mu\right)\left(\pi+\arctan\frac{\mu}{\Gamma}-\arctan\frac{\Delta+\mu}{\Gamma}\right)}\right].
\end{align}

\subsubsection{{\color{black}The} inter-band term in the case of $m_{1}^{*}=-m_{2}^{*}$ and $\mu \gtrsim \Delta$}
{\color{black}When the} Fermi energy is not close to the bottom of the conduction band, the logarithmic term {\color{black}can be} neglected and we obtain the most compact form of the inter-band spin-relaxation rate as
\begin{align}
\Gamma_{\textrm{s}}^{\textrm{inter}}=\frac{4\left|L\left(k_{\textrm{F}}\right)\right|^{2}\Gamma}{4\Gamma^{2}+\Delta^{2}\left(k_{\textrm{F}}\right)}.
\label{QuadIntra}
\end{align}

\subsubsection{{\color{black}Summary of the} result for the quadratic model}
Summing of intra- and inter-band term yields Eq.~(15) of the paper:
\begin{align}
\Gamma_{\textrm{s}}=\frac{4\left|\mathscr{L}\left(k_{\textrm{F}}\right)\right|^{2}\Gamma}{4\Gamma^{2}+\Delta_{\textrm{Z}}^{2}}+\frac{4\left|L\left(k_{\textrm{F}}\right)\right|^{2}\Gamma}{4\Gamma^{2}+\Delta^{2}\left(k_{\textrm{F}}\right)}.
\label{vegsogammas}
\end{align}

{\color{black}This is the main result, which is presented in the manuscript}.

We note that similar expressions can be obtained if the chemical potential lies in the nearby band.

\subsection{{\color{black}Linear band dispersion models}}

\begin{figure}
\hfill{}
\subfloat[]{
\includegraphics[width=0.45\columnwidth]{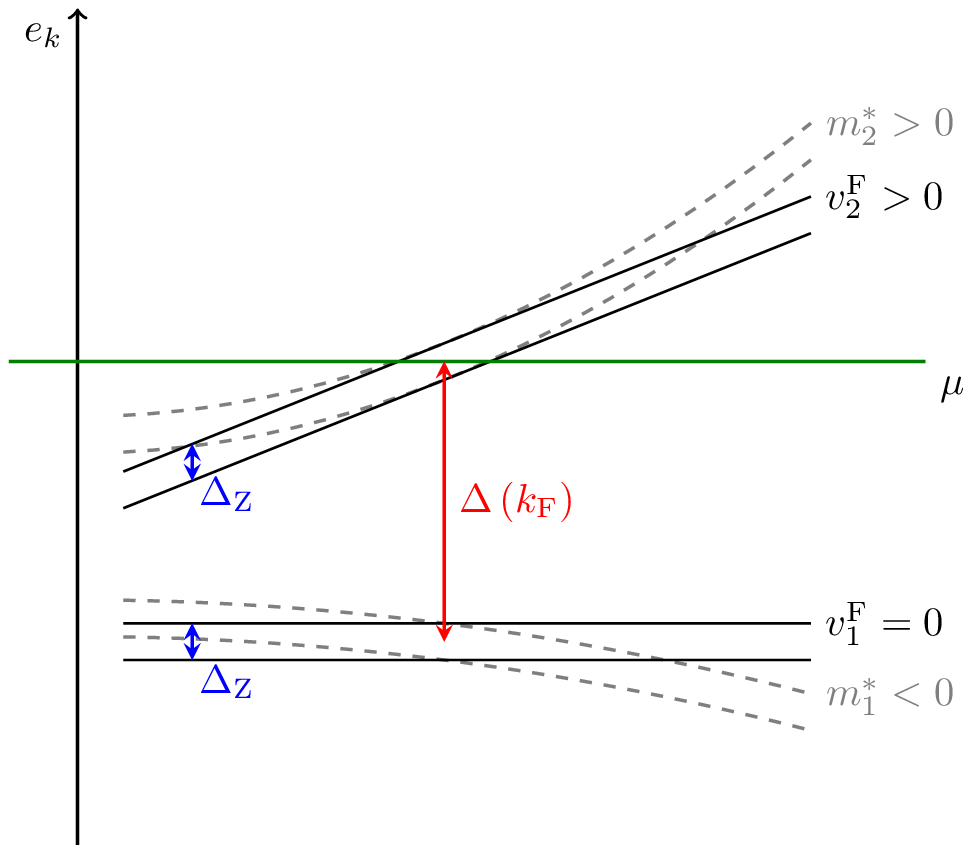}}
\hfill{}
\subfloat[]{
\includegraphics[width=0.45\columnwidth]{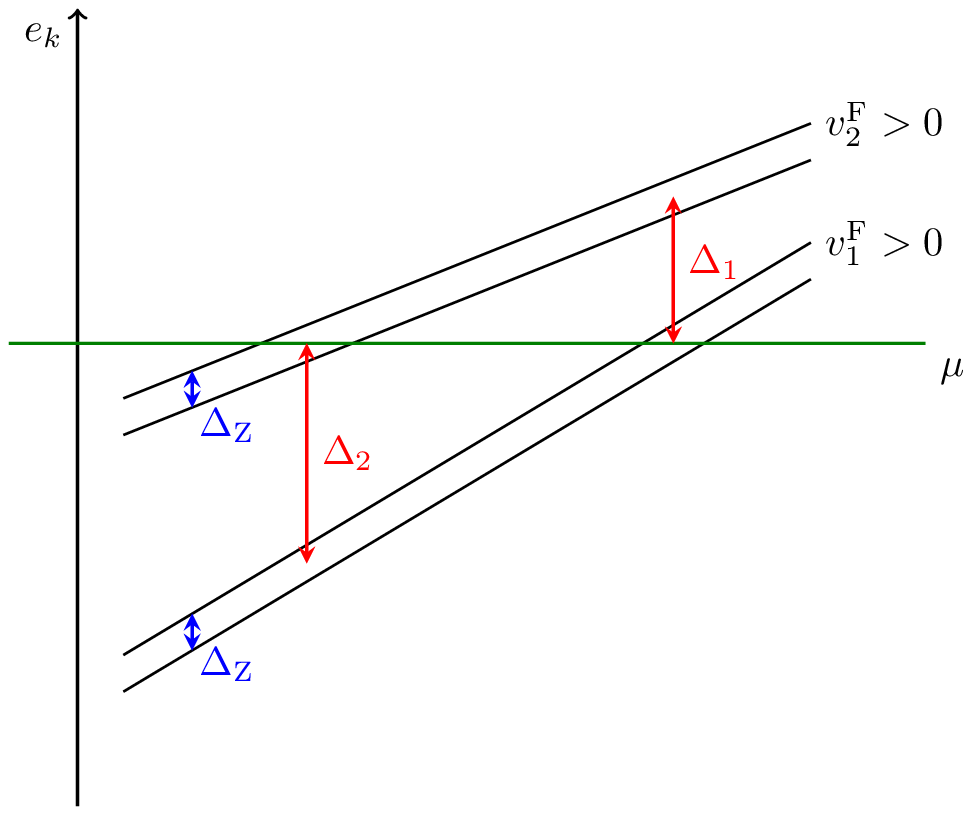}}
\hfill{}
\caption{{\color{black}Linear band dispersion models, with line slopes of the opposite (a) and the same sign (b). The first situation can be obtained e.g.~from linearizing a quadratic band dispersion in Fig.~\ref{QuadDisp} and the second occurs e.g.~for $\textrm{MgB}_{2}$ as shown in Ref. \onlinecite{SimonPRL2008}.}}
\label{LinearBandDispersion}
\end{figure}

{\color{black}Herein, we discuss the spin-relaxation time for linear model dispersions. The importance of studying this model is two-fold. First, every non-linear band dispersions can be linearized at the Fermi wave-vector and the plausible expectation is that the spin-relaxation rate can be obtained as a sum of the linearized segments. Second, spin-relaxation can be calculated for the linear band dispersion model and as we show below, the qualitative result, i.e.~dependence of $\Gamma_\textrm{s}$ on $\Gamma$ for the intra- and inter-band processes, is unchanged compared to the quadratic band dispersion even if the numerical factors are different. This proves that our calculation of the spin-relaxation is robust against the details of the band dispersion.

The linear band-dispersion models can take two characteristically different scenarios: those with lines with slopes of the opposite and the same sign. The situation is depicted in Fig.~\ref{LinearBandDispersion}. The first situation can be obtained e.g.~from linearizing a quadratic band dispersion in Fig.~\ref{QuadDisp} the Fermi wavenumber and the second occurs e.g. for $\textrm{MgB}_{2}$
 as shown in Ref. \cite{SimonPRL2008}.

\subsubsection{Linear band dispersions with the opposite slope}

We consider that the higher lying conduction band has a positive Fermi velocity thus $\epsilon_{k,2}^{\textrm{lin}}	 =\hbar v_{2}^{\textrm{F}} \left(k-k_{\textrm{F}}\right)+\mu$, where the Zeeman-energy is neglected. The nearby valence band can be approximated with a flat band with zero Fermi velocity: $\epsilon_{k,1}^{\textrm{lin}}=-\Delta+\mu m_{2}^{*}/m_{1}^{*}$, where the Zeeman splitting is also neglected too.

Then, our calculation yields for the intra-band contribution to the spin-relaxation rate:
\begin{align}
\Gamma_{\textrm{s}}^{\textrm{intra}}=\frac{4\Gamma\left|\mathscr{L}\left(k_{\textrm{F}}\right)\right|^{2}}{4\Gamma^{2}+\Delta_{\textrm{Z}}^{2}}.
\end{align}

Similarly, we obtain for the inter-band term:
\begin{align}
\Gamma_{\textrm{s}}^{\textrm{inter}}=\frac{4\left|L\left(k_{\textrm{F}}\right)\right|^{2}\Gamma}{\Gamma^{2}+\Delta^{2}\left(k_{\textrm{F}}\right)},
\end{align}
which look likes as if it was obtained from the quadratic model Eq.~(\ref{QuadIntra}) except the multiplication factor of the $\Gamma^{2}$ in the denominator. This is the result that we considered a zero Fermi velocity of the valence band.

\subsubsection{Linear band dispersions with the same slope}

The second linear model has two linear bands (apart from the spin) with positive Fermi velocities of different magnitudes. The two bands cross the Fermi level at two separate points. The band structure of this model is depicted in Fig.~\ref{LinearBandDispersion}b. This model describes the spectrum around the Fermi energy in e.g.~$\textrm{MgB}_{2}$ as it was shown in Ref. \onlinecite{SimonPRL2008}.

The intra-band term is similar to the previous results and it reads:
\begin{align}
\Gamma_{\textrm{s}}^{\textrm{intra}}=\frac{4\Gamma\left|\mathscr{L}\left(k_{\textrm{F}}\right)\right|^{2}}{4\Gamma^{2}+\Delta_{\textrm{Z}}^{2}}.
\end{align}

The inter-band term can be expressed as
\begin{align}
\Gamma_{\textrm{s}}^{\textrm{inter}}=\frac{4\Gamma\left|L\left(k_{\textrm{F}}\right)\right|^{2}}{\frac{{\left(\Delta_{1}+\Delta_{2}\right)}^{2}}{\Delta_{1}\Delta_{2}}\Gamma^{2}+\Delta_{1}\Delta_{2}},
\end{align}
where $ \Delta_{1}$ and $ \Delta_{2} $ are the distances of the two bands when one of the bands cross the Fermi level. The formula is symmetric in these two variables which means that the two bands change their roles as conduction and valence bands for the two Fermi level crossing points.

A special case is when $v_{1}^{\textrm{F}}=v_{2}^{\textrm{F}}$, i.e.~when the two linear bands are parallel therefore $\Delta_{1}=\Delta_{2}=\Delta$. This yields a result which is similar to the case of the quadratic dispersion and reads:
\begin{align}
\Gamma_{\textrm{s}}^{\textrm{inter}}=\frac{4\Gamma\left|L\left(k_{\textrm{F}}\right)\right|^{2}}{4\Gamma^{2}+\Delta^{2}}.
\end{align}
}

\section{The spin-relaxaton for a model with Rashba-like SOC}
Now we determine the spin-relaxation rate of a model where only Rashba-type SOC is present.
The Rashba-like SOC can be written as
\begin{align}
\mathcal{H}_{\textrm{SO}} =\sum_{k,\alpha}\hbar\lambda\left(\sigma_{x}k_{y}-\sigma_{y}k_{x}\right) ,
\end{align}
where $\sigma_{x}, \sigma_{y}$ are the Pauli matrices. The matrix element of intra-band SOC can be expressed as $\left|\mathscr{L}\left(k\right)\right|=\hbar\lambda k$. We can expand Eq.~(\ref{vegsogammas}) to get
\begin{align}
\Gamma_{\textrm{s}}=\frac{8m_{2}^{*}\mu\lambda^{2}}{4\Gamma^{2}+\Delta_{\textrm{Z}}^{2}}.
\end{align}
Using the spin and momentum life-times instead of relaxation-rates, we obtain
\begin{align}
\frac{1}{\tau_{\textrm{s}}} & =\frac{8m_{2}^{*}\mu\lambda^{2}}{\hbar^{2}}\frac{\tau}{1+\left(\frac{\Delta_{\textrm{Z}}\tau}{\hbar}\right)^{2}}.
\end{align}
A similar expression was obtained recently (Eq.~(40) in \cite{BalentsBurkovPRB2004}) for this particular case.
\end{widetext}



\begin{thebibliography}{10}
\bibitem{Endnote}
\bibinfo{note}{Magnetic field, however splits the energy degeneracy.}
\end{thebibliography}

\end{document}